\def\year{2022}\relax
\documentclass[letterpaper]{article} 
\usepackage{aaai22}  
\usepackage{times}  
\usepackage{helvet}  
\usepackage{courier}  
\usepackage[hyphens]{url}  
\usepackage{graphicx} 
\usepackage{natbib}  
\usepackage{caption} 
\DeclareCaptionStyle{ruled}{labelfont=normalfont,labelsep=colon,strut=off} 
\usepackage{algorithm}
\usepackage{algpseudocode}
\usepackage[export]{adjustbox} 
\usepackage{subfigure}
\usepackage{epstopdf}
\usepackage{booktabs}
\usepackage{amssymb}
\usepackage{amsmath}
\usepackage{multirow}

\usepackage{newfloat}
\usepackage{listings}
\floatstyle{ruled}
\newfloat{listing}{tb}{lst}{}
\floatname{listing}{Listing}
\title{Re-ranking With Constraints on Diversified Exposures for Homepage Recommender System}
\author{
    Qi Hao\textsuperscript{\rm 1},Tianze Luo\textsuperscript{\rm 2},Guangda Huzhang\textsuperscript{\rm 1}
}
\affiliations{
    \textsuperscript{\rm 1}Alibaba Group,Hangzhou,China, \textsuperscript{\rm 2}Nanyang Technological University,Singapore\\

    \textsuperscript{\rm1}{qihao.data@gmail.com,guangda.hzgd@alibaba-inc.com},\textsuperscript{\rm2}tianze001@e.ntu.edu.sg\\
}

\usepackage{bibentry}

\begin{document}

\maketitle

\begin{abstract}
The homepage recommendation on most E-commerce applications places items in a hierarchical manner, where different channels display items in different styles. Existing algorithms usually optimize the performance of a single channel. So designing the model to achieve the optimal recommendation list which maximize the Click-Through Rate (CTR) of whole homepage is a challenge problem. Other than the accuracy objective, display diversity on the homepage is also important since homogeneous display usually hurts user experience. In this paper, we propose a two-stage architecture of the homepage recommendation system.
In the first stage, we develop efficient algorithms for recommending items to proper channels while maintaining diversity. The two methods can be combined: user-channel-item predictive model with diversity constraint. In the second stage, we provide an ordered list of items in each channel. Existing re-ranking models are hard to describe the mutual influence between items in both intra-channel and inter-channel. Therefore, we propose a Deep \& Hierarchical Attention Network Re-ranking (DHANR) model for homepage recommender systems. The Hierarchical Attention Network consists of an item encoder, an item-level attention layer, a channel encoder and a channel-level attention layer. Our method achieves a significant improvement in terms of precision, intra-list average distance(ILAD) and channel-wise Precision@k in offline experiments and in terms of CTR and ILAD in our online systems.
\end{abstract}

\section{INTRODUCTION}
Homepage recommendation is a common task in the industry, especially on E-commerce platforms, such as Amazon and Aliexpress. In many cases, the homepage has a hierarchical architecture containing multiple channels, and each channel contains several items, as shown in figure 1. A channel, such as Top Sellers and Deal of the Day, is a collection of items gathered according to certain attributes of items. Items can be recommended to multiple channels according to their corresponding attributes. Existing algorithms only consider the accuracy and diversity of a single channel and are hard to produce the optimal recommendation list of the whole homepage. Thus designing a model for optimal recommendation on the whole list is a challenge problem. 
\begin{figure}
  \centering
  \includegraphics[width=0.76\linewidth]{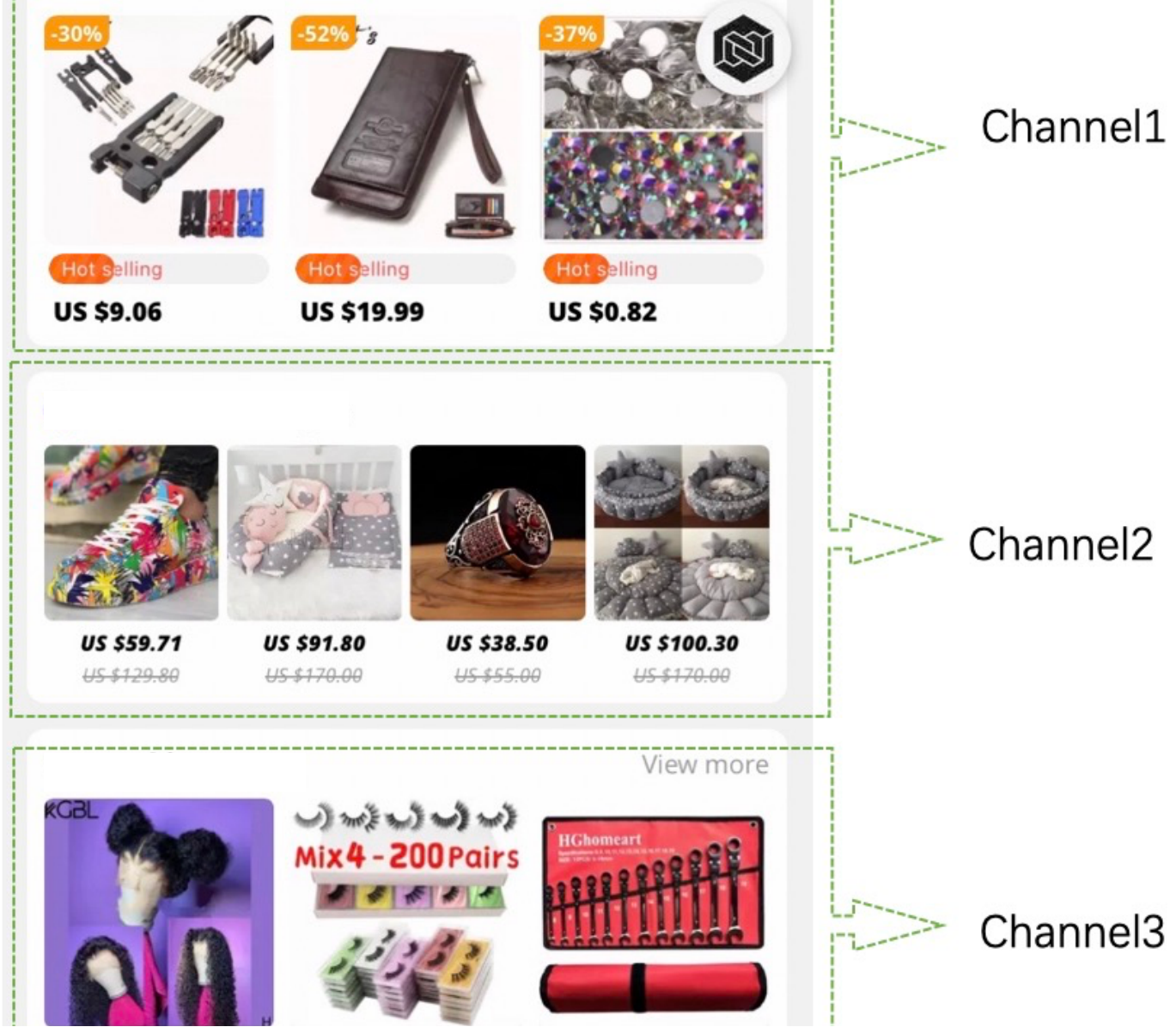}
  \caption{The hierarchical structure of homepage.}
  \label{fig:homepage}
\end{figure}

In addition to the recommendation accuracy, the differences of users’ preferences for channels and the diversity of the whole homepage have an important impact on homepage recommendation. Firstly, we discuss whether users have different preferences for channel. Based on behavioral data, such as the frequencies of browsing in each channel, we classify users into three groups. As expected, users can be clustered into three different groups based on their different channel preferences. Therefore, instead of user–item predictive model, we require to create a user–channel–item predictive model for CTR prediction. Besides, in order to design an accurate recommender for CTR prediction, modeling the mutual influence between items is an important topic for the researchers of E-commerce recommender systems. Secondly, diversification has become one of the leading topics of recommender system research. We find it is necessary to manage diversity in homepage recommender systems. Furthermore, we discuss whether users have different levels of tolerance for item repetitions belonging to each category. We calculate the CTR under different number of phones', clothes', food's and jewelry's recurring during a month, as shown in figure 2, which turns out that users' tolerance for reappearance of different category is different. For example, the CTR is the highest when the number of phones' recurring equals to 2, however, it is the highest when the number of clothes' recurring equals to 3, so we suggest to set specific threshold for each category.

Existing methods are sub-optimal because it is hard to tackle the above issues. Although several re-ranking models have been proposed to model the mutual influence between items, they do not specifically consider the homepage structure, which usually exhibits a hierarchical architecture. Also, existing re-ranking methods do not consider avoiding repeated appearances of the same item in multiple channels. Although there exists abundant methods to deal with diversification in ranking, they do not fit well with homepage recommendation applications, because it is hard to balance accuracy and diversity given multiple channels. Besides, these methods are not as adequate for modeling the relationship between items as Deep List-wise Context Models. They only use pairwise dissimilarities to characterize the overall diversity property of the list, which may not capture complex relationships among items. For example, in maximal marginal relevance (MMR) \cite{carbonell1998use} method , the score of an item under consideration is proportional to its relevance minus a penalty term that measures its similarity to the items previously selected.

\begin{figure}
  \centering
  \includegraphics[width=0.85\linewidth]{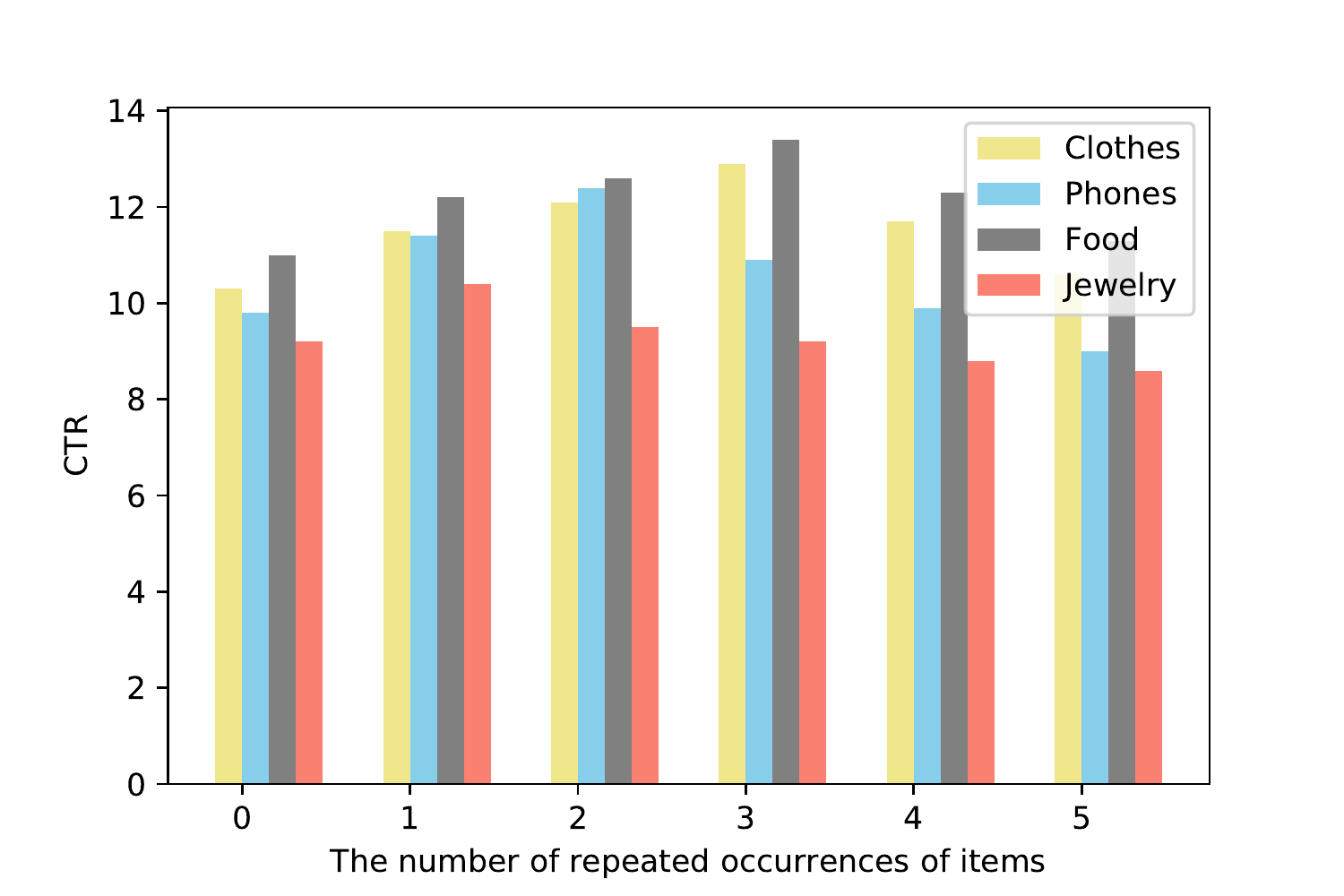}
  \caption{Distributions of CTR under different number of phones', clothes', food's and jewelry's reappearances.}
  \label{fig:tolerance}
\end{figure}

\begin{figure*}[t]
\centering
\includegraphics[width=17.5cm]{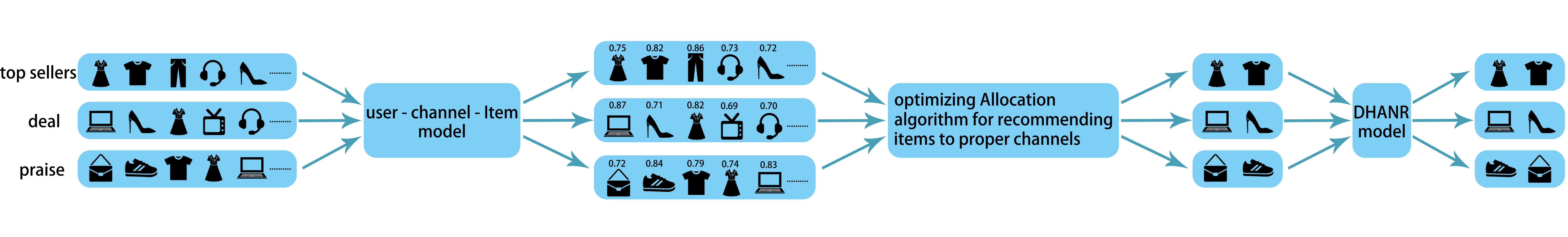}
\caption{Method Overview}
\label{fig:overflow}
\end{figure*}

To summarize, the main contributions made in this paper are as follows:
\begin{itemize}
\item We propose an efficient two-stage architecture of homepage recommender. To the best of our knowledge, it is the first to explicitly introduce the homepage structure into re-ranking task in large-scale online system. 
\item We design a user-item-channel predictive model for CTR prediction as well as diversified assignment. Specifically, we propose a general framework that employs linear programming to recommend items to proper channels under a whole class of diversity constraints.
\item We provide an optimal recommendation list in each channel by re-ranking. In order to describe the mutual influence between items in both intra-channel and inter-channel, we propose a Deep \& Hierarchical Attention Network Re-ranking (DHANR) model for homepage recommender systems. 
\item Our proposed method can be specific to determine how many items should be recommended for each category. It is better to set a specific threshold for each category, as the tolerance of the user to the multiple occurred items in each channel is different.
\item We conduct extensive offline and online experiments on an E-commerce Recommendation System. We observe a superior performance of our proposed model.
\end{itemize}

\section{RELATED WORK}
First we review related work on various re-ranking approaches, and then discuss existing measures of diversity in the recommender systems.

\textbf{Re-ranking Methods}.
Typically, a ranking function is learned from the labeled dataset to optimize the global performance, which produces accurate ranking score or maximum CTR \cite{burges2005learning,burges2010ranknet,cao2007learning,friedman2001greedy,joachims2006training,taylor2008softrank,xia2008listwise}. However these methods do not explicitly consider the mutual influence between items. Several deep list-wise context models \cite{ai2018learning,yin2016ranking,zhuang2018globally,pei2019personalized,bello2018seq2slate} have been proposed to model the mutual influence between items. These works tend to model the mutual influences between items explicitly to refine the initial list given by the previous ranking algorithm, which is known as re-ranking. The main objective is to build the scoring function by encoding the information of all items in the list into feature space. The methods for encoding the feature vectors contain RNN-based and transformer-based\cite{kang2018self}. The RNN-based models include GlobalRerank \cite{zhuang2018globally}, Seq2Slate \cite{bello2018seq2slate} and DLCM \cite{ai2018learning}, and the transformer-based model including PRM \cite{pei2019personalized}. They feed the initial list into RNN-based or transformer-based structure sequentially and output the encoded vector at each time step, to model the mutual influences between items. Seq2Slate \cite{bello2018seq2slate} and GlobalRerank \cite{zhuang2018globally} use the decoder structure to generate the re-ranked list. Seq2Slate \cite{bello2018seq2slate} uses the pointer network\cite{vinyals2015pointer} to generate re-ranked list sequentially. GlobalRerank \cite{zhuang2018globally} uses RNN with the attention mechanism\cite{vaswani2017attention} as the decoder. PRM \cite{pei2019personalized} directly optimizes the whole recommendation list by employing a transformer structure to efficiently encode the information of all items in the list.

\textbf{Diversity Methods}.
A common specific definition of diversity in the literature is the
average pairwise dissimilarity between recommended items. There are a wide range of methods for managing diversity in ranking\cite{carbonell1998use,borodin2012max,qin2013promoting,chen2018fast,wilhelm2018practical,hijikata2009discovery,lathia2010temporal,mcnee2006being,vargas2014coverage,yu2009takes,zhang2008avoiding,ziegler2005improving}. Most methods define set-wise diversity metrics and involve a tunable parameter to adjust the trade-off between relevance and diversity, such as maximal marginal relevance (MMR)\cite{carbonell1998use}, max-sum diversification (MSD) \cite{borodin2012max} and entropy regularizer (Entropy) \cite{qin2013promoting}. The maximal marginal relevance (MRR)\cite{carbonell1998use} model was one of the pioneering
work for promoting diversity in information retrieval tasks.
The trust-region based optimization method\cite{yuan2000review} aims to maximize the diversity of recommendation list, while maintaining an acceptable level of matching quality. Entropy regularizer (Entropy) \cite{qin2013promoting} method is incorporated in the contextual combinatorial bandit framework to diversify the online recommendation results. 
There are also methods using a DPP-based \cite{borodin2009determinantal,gillenwater2014approximate,kulesza2012determinantal} approach for recommending relevant and diverse items to users. For example, a fast greedy maximum a posteriori(MAP) inference algorithm for determinantal point process \cite{chen2018fast} was proposed to improve recommendation diversity. 

\section{METHOD}
We propose a two-stage architecture for the homepage recommender. The overview of our method is shown in figure 3. The first stage includes a recommendation process that puts items into proper channels under diversity constraints. In the second stage, we propose the DHANR model to provide an ordered list of items in each channel. 

\subsection{Diversity-Constrained Channel Items Recommender Framework}
Personalized recommendation with diversity considerations is an important task for modern recommender systems. Our framework deals with this problem.

\textbf{A user–channel–item Predictive Model For CTR Prediction}.
In order to model the differences of users’ preferences for channels, we create a user–channel–item predictive model to predict the CTR in each channel. Typically, ranking in recommender system only considers the user-item pair features. In our deep network, we also consider the channel features, which are used as the input for the context network. Thus, the outputs of the model are user–channel–item tripartite structure scores. Popular model structures, such as DNN, DeepFM, DCN, can be used here.

\begin{figure*}[t]
\centering
\includegraphics[width=18cm]{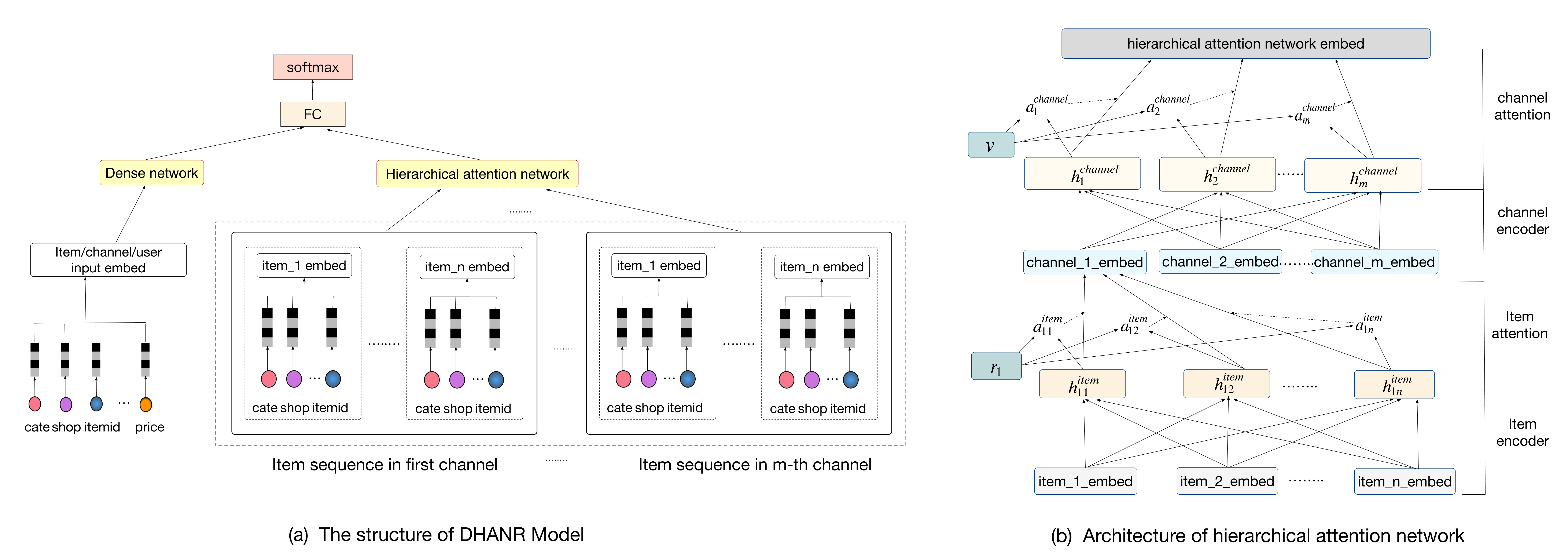}
\caption{The structure of Deep \& Hierarchical Attention Network Re-ranking (DHANR) Model}
\label{fig:position_model3}
\end{figure*}

\textbf{Diversity Measure}.
We choose to define diversity with respect to attributes of items, such as category, brand and style. How to set the thresholds for each type of recommended item requires further consideration.  For example, how many items related with food can be displayed on the homepage. We calculate the conditional probability of CTR given the number of items belong to category $c$ equals to $k$ as:
\begin{equation}    \label{eq}
CTR_{\text{category}}=\frac{Pr[\text{click}=1,cnt(\text{category}=c)=k]}{Pr[cnt(\text{category}=c)=k]}  \\
\end{equation}

where $cnt(\text{category}=c))$ means the number of the items belong to category $c$. The $k$ maximizes the above formula is expressed by $T_c$. This $T_c$ means the maximum of items belonging to category $c$ can be displayed on homepage during one exposure.  

\textbf{Optimizing Allocation with Linear Programming}.
We model the problem to find the optimal CTR of the whole page, under the diversity constraints, as a linear programming problem. The goal is to obtain the maximal CTR of the whole homepage while satisfying all diversity constraints, which can be formulated as the following,

\begingroup\makeatletter\def\f@size{8}\check@mathfonts
\begin{equation}
\begin{split}  
&\max_{X\in\Omega} \quad \sum_{i=1}^{M} \sum_{j=1}^{N} R_{ij}X_{ij}-U\sum_{c=1}^{S} \xi_c  \\
&s.t.\quad  \left\{\begin{array}{lc}  
\sum_{i=1}^{M} X_{ij}\leq 1,\forall j\in [N]\\ [2mm]
\sum_{j=1}^{N} X_{ij} = V_i+h,\forall i \in [M]\\ [2mm]
\sum_{i=1}^{M} \sum_{j\in K_c} X_{ij}\leq T_c+\xi_c,\forall c \in [S]\\[2mm]
\sum_{j\in K_c} X_{ij}\leq B,\forall c \in [S], i \in [M]\\[2mm]
X_{ij}\in \left \{0,1\right \} ,\forall j\in [N],i \in [M]\\ [2mm]
\xi_c \geq 0,\forall c\in [S]
\end{array}\right.  
\end{split}
\end{equation}
where $ \Omega =  \left \{\mathbb{R^{M\times N}} \right \} $ , $ X_{ij} $ is the decision variable indicating whether the algorithm allocates the $j$-th item to the $i$-th channel. $R_{ij}$ is the CTR of item $j$ in channel $i$, which is calculated by user–channel–item Predictive Model. The first restriction means that each item only can be assigned to at most one channel. The second restriction ensures that $V_i +h$ items must be allocated to the $i$-th channel. $V_i$ is the number of recommended items in the $i$-th channel. $h$ is the hyperparameter which controls the number of items input into re-ranking models. The third restriction is the constraints on diversified exposures, which ensures that the number of items belong to category $c$ displayed on the homepage should not exceed the diversity constraint threshold $T_c$. For example, the items related with food cannot be displayed on homepage more than 3 positions. Besides, we introduce a slack variable $\xi_{c}$ to control the strength of diversity constraints. The fourth restriction ensures the number of items belonging to the same category displayed on each channel should not exceed the bound $B$.

The resulting linear program can be solved efficiently and optimally with standard algorithms like interior point methods. Then we allocate items to the channels accordingly. The output item lists contain $V_i+h$ items in each channel.

\subsection{Deep \& Hierarchical Attention Network Re-ranking (DHANR) Model}

In this section, we create a re-ranking model to optimize the order of items in each channel. In order to model the mutual influence between items in both intra-channel and inter-channel as well as the differences of users’ preferences, we build a Deep \& Hierarchical Attention Network Re-ranking (DHANR) model to predict the CTR in each channel. The optimal recommendation list in each channel can be obtained by ranking items by their re-ranking scores.

The architecture of our DHANR Model is shown in figure 4 (a). The model consists of two categories of components: the deep component and the hierarchical attention component. The deep component has been introduced in user-channel-item model. The detailed structure of the hierarchical attention network component is shown in figure 4 (b). The input of this network is the item lists generated by previous user-channel-item and allocation method. Our hierarchical attention network consists of an item encoder, an item-level attention layer, a channel encoder and a channel-level attention layer.

\textbf{Item encoder}.
Given a sequence of items in each channel, we embed the items vectors, then adopting Transformer-like encoder to integrate the mutual influences of item-pairs in each channel. The self-attention mechanism is suitable in our re-ranking task as it directly models the mutual influences for any two items regardless the distances between them. The self-attention function is defined as:
\begin{equation}  
Attention(\boldsymbol Q,\boldsymbol K,\boldsymbol V) = softmax(\frac{\boldsymbol Q {\boldsymbol K}^T}{\sqrt{d_k}})\boldsymbol V,
\end{equation}  

where $\boldsymbol Q,\boldsymbol K,\boldsymbol V$ are matrices and represent queries, keys and values respectively. $d_k$ is a scaling factor and represent the dimensionality of matrix $\boldsymbol k$. To model more complex mutual influences, we use the multihead attention. Our encoding module consists of $N_b$ blocks of Transformer encoder. Each block contains an self-attention layer
and a Feed-Forward Network (FFN) layer. We use the aforementioned encoding method to generate a richer representation of items in each channel, that is $h^{\text{im}}_{ij}$.

\textbf{Item-level attention layer}.
We introduce the attention mechanism and an item level context vector $r_i$ to extract items that are important to the channel and aggregate the representations of those informative items to form a channel vector. 
\begin{equation}  
u^{\text{im}}_{ij}=tanh(W_w h^{\text{im}}_{ij}+b_w);a^{\text{im}}_{ij}=\frac{exp({u^{\text{im}}_{ij}}^T r_i)}{\sum_j exp({u^{\text{im}}_{ij}}^T r_i)} ,
\end{equation} 
\begin{equation}  
s_i=\sum_j a^{\text{im}}_{ij} h^{\text{im}}_{ij} , \\
\end{equation}

\textbf{Channel encoder}.
Given the channel vectors $s_i,i \in [M]$, we adopt an encoder as illustrated in the item encoder to integrate the mutual influences of channel-pairs in a similar way as the item encoder and obtain a richer representation of channels, $h^{\text{cha}}_i$. 

\textbf{Channel-level attention layer}.
We again use the attention mechanism and introduce a channel level context vector $v$ to extract overall homepage representation.
\begin{equation}  
u^{\text{cha}}_i=tanh(W_s h^{\text{cha}}_i+b_s);a^{\text{cha}}_i=\frac{exp({u^{\text{cha}}_i}^T v)}{\sum_i exp({u^{\text{cha}}_i}^T v)} ,\\
\end{equation} 
\begin{equation}  
t=\sum_i a^{\text{cha}}_i h^{\text{cha}}_i ,\\
\end{equation} 

Finally, the vectors generated by the deep component and by the hierarchical attention component are concatenated together to feed into a multilayer perceptron (MLP).

\section{EXPERIMENTAL RESULTS}
In this section, we first introduce the datasets and baselines used for evaluation. Then we compare our methods with baselines on these
datasets to evaluate the effectiveness of our model. 

\subsection{Datasets}
To the best of our knowledge, there does not exist publicly available re-ranking homepage datasets with context information for recommendation. Therefore, we construct an E-commerce Re-ranking dataset from a real-world E-commerce platform. The dataset contains the user click-through data of the homepage. The size of training and testing set is about 320 millions and 80 millions respectively. These channels have 3, 4, and 3 items, and the number of items is about 8M. Compared with channel 3, channel 1 and channel 2 have 1.07\% and 1.14\% higher CTR.


\subsection{Baselines}
As there are two stages in our method, we conduct the study to understand the contributions of each stage. In the first stage, the algorithm for CTR prediction and recommending items to proper channels under diversity constraints is denoted as UCI-AA. We use following methods as our baselines in the first stage. 

\textbf{DNN}: The concatenation of all the embedding vectors is
fed into multiple layer perception. Then, an output layer
predicts the probability whether the user will click the item
with a sigmoid function.

\textbf{DNN(single)}: We also train single-channel models by training a separate model for each channel.

\textbf{MMR}: Maximal marginal relevance is proposed
to reduce redundancy while maintaining relevance.

\textbf{MSD}: The objective function comprises of two terms, a modular relevance term and a supermodular sum of distance diversity term.

\textbf{UCI-AA}: The model proposed in stage 1 of this paper, which recommends items to proper channels under diversity constraints based on CTR score as predicted in user-channel-item model. The UCI-AA method determines the set of items under each channel, but it has not determined the order of items in each channel. In the second stage, we add the hierarchical attention network to re-ranking, which is denoted as UCI-AA-DHANR. We choose DLCM\cite{ai2018learning} and PRM \cite{pei2019personalized} as our baseline methods in the second stage, as other re-ranking methods mentioned in related work can not be parallelized in online inference and the time complexities are unacceptable.

\textbf{UCI-AA-DLCM}: It is a re-ranking model. The GRU is used to encode the local context information into a global vector. Combing the global vector and each feature vector, it learns a more powerful scoring function than the model without re-ranking.

\textbf{UCI-AA-PRM}: It directly optimizes the whole recommendation
list by employing a Transformer structure to efficiently encode the information of all items in the list.

\textbf{UCI-AA-DHANR}: The model proposed in this paper. Based on the UCI-AA model, we propose a Deep \& Hierarchical Attention Network Re-ranking (DHANR) model for describing the mutual influence between items in both intra-channel and inter-channel.

\subsection{Evaluation Metrics}
For offline evaluation, we use precision to compare different methods for the homepage recommendation: 
\begin{eqnarray}  
Precision=\frac{1}{|U|} \sum_{u \in U} \frac{\left|R_u \cap C_u\right|}{\left|R_u\right|},
\end{eqnarray}

where $U$ is the set of all user requests in the dataset, $R_u$ is the set of items recommended to user request $u$, $C_u$ denotes the set of items clicked. Furthermore, in order to compare the performance of our DHANR Model with other re-ranking methods, we use ${Precision@k}_{channel}$ to evaluation the performance in each channel, which is defined as the the fraction of clicked items in the top-k recommended items for all test samples in each channel. Since we are optimizing the recommendation of multiple channels as a whole, $Precision@k$ is not suitable for the homepage recommendation. 
\begin{eqnarray}  
{Precision@k}_{channel}=\frac{1}{|U|} \sum_{u \in U} \frac{\sum_{i=1}^k \mathbb I (S_u(i)) }{k},
\end{eqnarray}
where $S_u$ is the ordered list of items given by the re-ranking model for each request $u \in U$ and $S_u(i)$ is the $i$-th item. $\mathbb I$ is the indicator function on whether item $i$ is clicked or not.

Also, we compare the diversity of our method with other diversity methods. Diversity metric is measured by the intra-list average distance (ILAD), which defined as follows:
\begin{eqnarray}  
Diversity=avg_{u \in U}avg_{i,j \in R_u,i \neq j} (1- \boldsymbol S_{ij}),
\end{eqnarray} 
where $\boldsymbol S_{ij}$ denotes the similarity between two items $i$ and $j$, for fair comparison, we define $\boldsymbol S_{ij}$ by the Jaccard similarity between the categories of two items $i$ and $j$.

For online A/B tests, we use CTR and diversity as metrics. CTR is the click through rate and can be calculated by IPV/PV. PV and IPV are defined as the total number of items viewed and clicked by the users, respectively.  

\subsection{Experimental Settings}
In the experiments, for both baselines and our methods, we use the same value for those critical hyper parameters. The learning rate of Adam optimizer is 0.0001. $p_{\text{dropout}}$ is set to 0.01. The batch size is set to 512. For the deep network, we set the hidden layers as 3, the number of units in each hidden layer is set to 512, 256, 128 respectively, and we use ReLUs as the activation function. The rest of the settings belonging to the customized parts of our model will be listed at the corresponding parts in the
Hyper-Parameters Investigation section. The experiments are repeated 10 times with independent data. 80\% as training data, 20\% as test data. We display the mean performance in the following tables in offline and online performance sections. The standard deviation is within the range of  $[0.002, 0.009]$ .

\begin{table*}[t]
\caption{Offline Precision \& Diversity of Methods during a week.}
  \label{tab:freq}
\centering
\begin{tabular}{|c|c|c|c|c|c|c|c|c|}
\hline

\multirow{2}{*}{method}& \multicolumn{4}{c|}{Precision(\%)} &\multicolumn{4}{c|}{Diversity(\%)}\\
\cline{2-9}
{}&total&channel 1&channel 2&channel 3&total&channel 1&channel 2&channel 3\\
\hline
DNN(single)&11.86&12.15&13.39&11.03&46.62&47.23&46.81&46.25\\
\cline{1-9}
DNN&12.84&12.76&13.56&11.85&50.54&52.42&50.76&50.43\\
\cline{1-9}
MMR&13.29&13.20&13.83&12.16&58.71&58.92&59.14&58.13\\
\cline{1-9}
MSD&13.25&13.24&13.88&12.18&57.28&57.43&57.65&56.62\\
\cline{1-9}
UCI-AA&14.01&13.95&15.22&12.82&65.84&66.15&66.27&65.14\\
\cline{1-9}
UCI-AA-DLCM&14.28&14.14&15.34&13.04&65.83&66.17&66.25&65.18\\
\cline{1-9}
UCI-AA-PRM&14.32&14.21&15.36&13.05&65.86&66.15&66.28&65.17\\
\cline{1-9}
UCI-AA-DHANR&14.84&14.75&15.82&13.62&65.89&66.18&66.31&65.19\\

\hline
\end{tabular}
\end{table*}

\subsection{Offline Performance}
In this section, we conduct offline evaluations on an E-commerce dataset. We conduct the study to find how each stage in our model contributes to the performance. In the first stage, the percent difference in precision and diversity for the various approaches during a week is described in table 1. In the second stage, we illustrate the Precision@1, Precision@2 of our DHANR model and other re-ranking methods in each channel, as listed in table 3. We also conduct the Hyper-Parameters Investigation.
\subsubsection{\textbf{Model analysis}}
Table 1 shows that our UCI-AA model achieves stable and significant performance improvements comparing with all baselines. UCI-AA outperforms MSD by 5.7\% at Precision and
15.1\% at Diversity, as well as outperforms MMR by 5.4\% at Precision and 12.5\% at Diversity. The gap gets larger when comparing with DNN, which has 9.3\% increase at Precision and 25.6\% increase at Diversity. DNN(single) achieves the worst performance due to training single-scene models does not take advantage of the rich user and item information from all channels.
Among all the compared diversity methods, our proposed UCI-AA almost always achieve the best performance in every channel, which mainly comes from the powerful encoding of user' preferences for channels and the effective controlling of diversified exposures. In general, we see that the model tends to demote listings with more diversity. The diversity score distribution of the categories and brands in figure 5 is quite revealing. The plot clearly shows our method frequently produce recommendations with different categories and brands.

Further more, with re-ranking method, UCI-AA-DHANR extend 5.6\% at Precision over UCI-AA. table 2 shows that on average, UCI-AA-DHANR achieves 11.5\% improvements on Precision@1 and 6.2\% improvements on Precision@2 comparing with UCI-AA due to the powerful encoding of mutual influence between items in both intra-channel and inter-channel. Among all the compared reranking methods, our proposed DHANR almost always achieve the best performance in every channel. For further illustration, in order to validate that the DHANR model is able to learn meaningful information with respect to characteristics of items and channels, we visualize the average attention weights between items on category. As shown in figure 6, the items with similar categories tend to have larger attention weights. We can conclude that the DHANR model can successfully capture mutual-influences of items. For example, “men’s clothing” has more influences on “women’s clothing” than on “consumer electronics”. Besides, the channels with similar items tend to have larger attention weights.

\subsubsection{\textbf{Hyper-Parameters Investigation}}
In UCI-AA model, we test the impact of trade-off parameter $U$ in expression (3). The results are shown in figure 7. As $U$ increases, the diversity constraint become stronger. Precision improves at first, achieves the best value when $U \approx 2$, and then decreases a little bit. Diversity is monotonously increasing as $U$ increases. Therefore, we set $U$ as 2. In addition to the trade-off parameter $U$, we also try different bounds($B=1; 2; 3$) which restrict the number of items from the same category that can be displayed on each channel and different hyperparameters ($h=0; 1; 2; 3$). Due to limited space, we will not put the experimental data. The best performance is achieved when $B$ is 2 and $h$ is 1. In DHANR model, we try different settings($h=1; 2; 3$) in the multi-head attention layer and different settings of block number($N_b=3; 4; 5$). No significant improvements are observed in the DHANR model, which are listed in table 2. 

\begin{figure}[t]
\centering
\subfigure[Category distance.]{
\begin{minipage}[t]{0.5\linewidth}
\centering
\includegraphics[width=4cm]{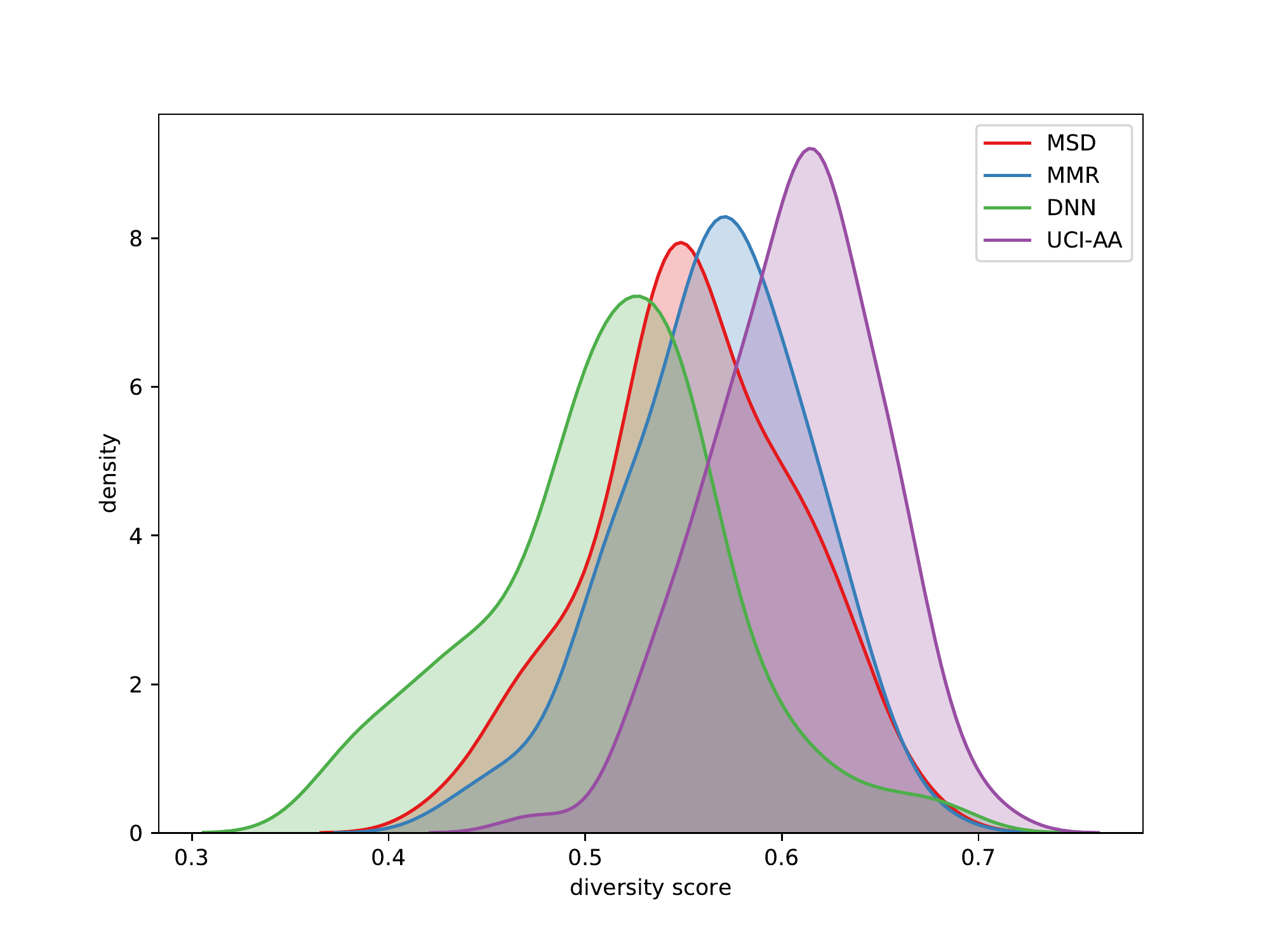}
\end{minipage}%
}%
\subfigure[Brand distance.]{
\begin{minipage}[t]{0.5\linewidth}
\centering
\includegraphics[width=4cm]{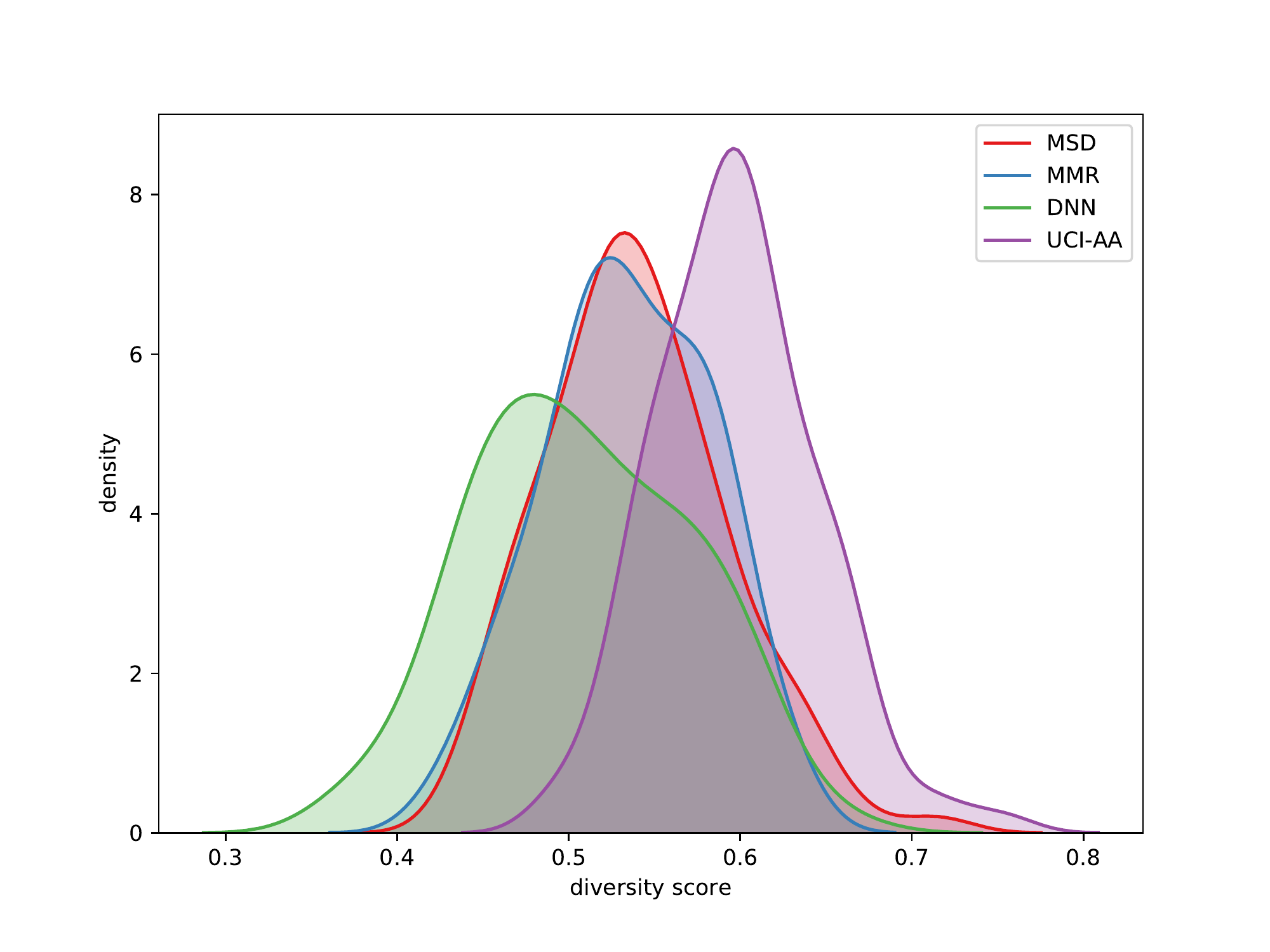}
\end{minipage}%
}%
\centering
\caption{Diversity score distribution of the recommender systems, based on the similarity of (a) category, (b) brand}
\label{fig:category}
\end{figure}

\begin{figure}
  \centering
  \includegraphics[width=0.9\linewidth]{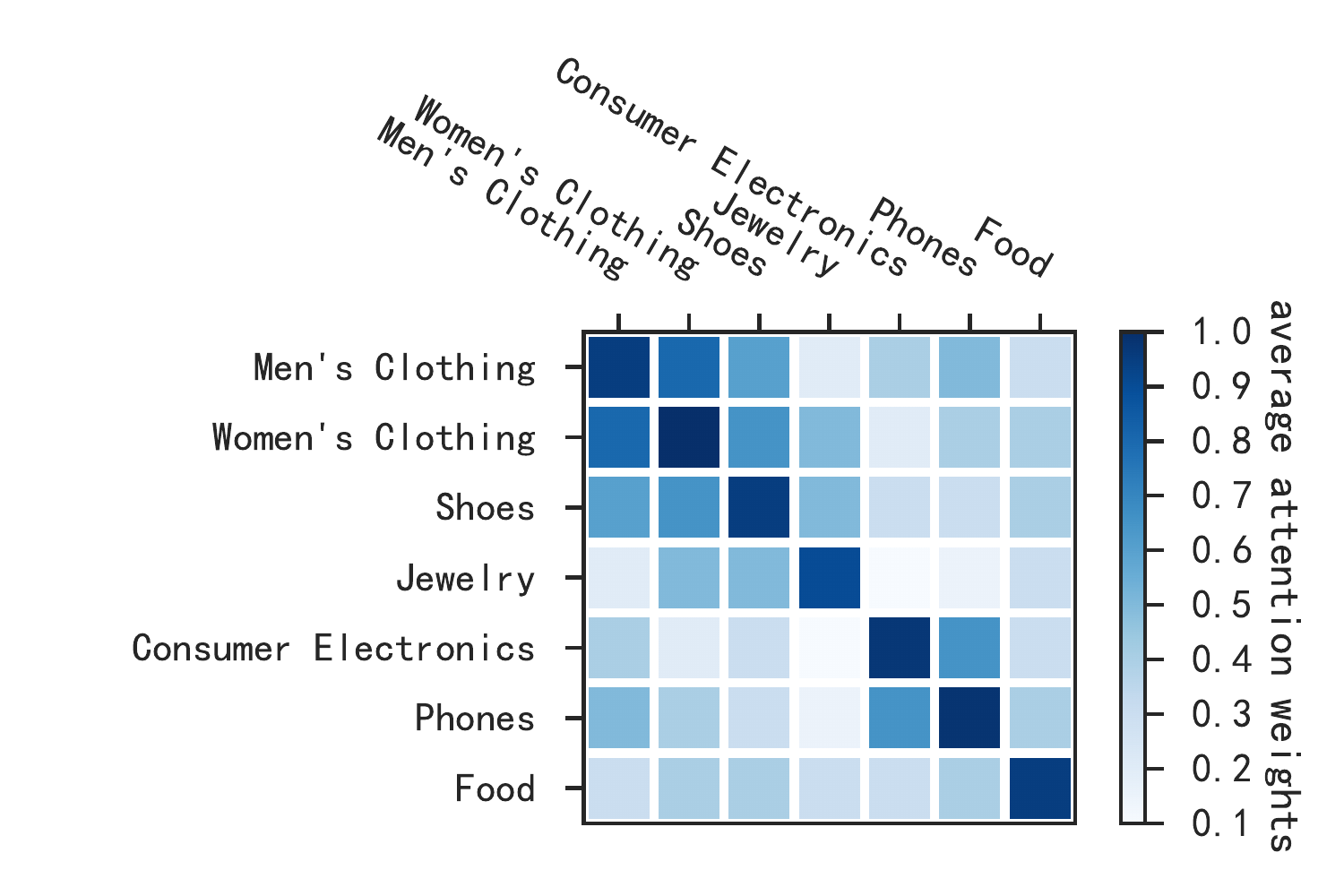}
  \caption{Average attention weights related to items’ attributes.}
  \label{fig:attention1}
\end{figure}

\begin{figure}[t]
\centering
\subfigure[impact of $U$ on precision.]{
\begin{minipage}[t]{0.5\linewidth}
\centering
\includegraphics[width=4cm]{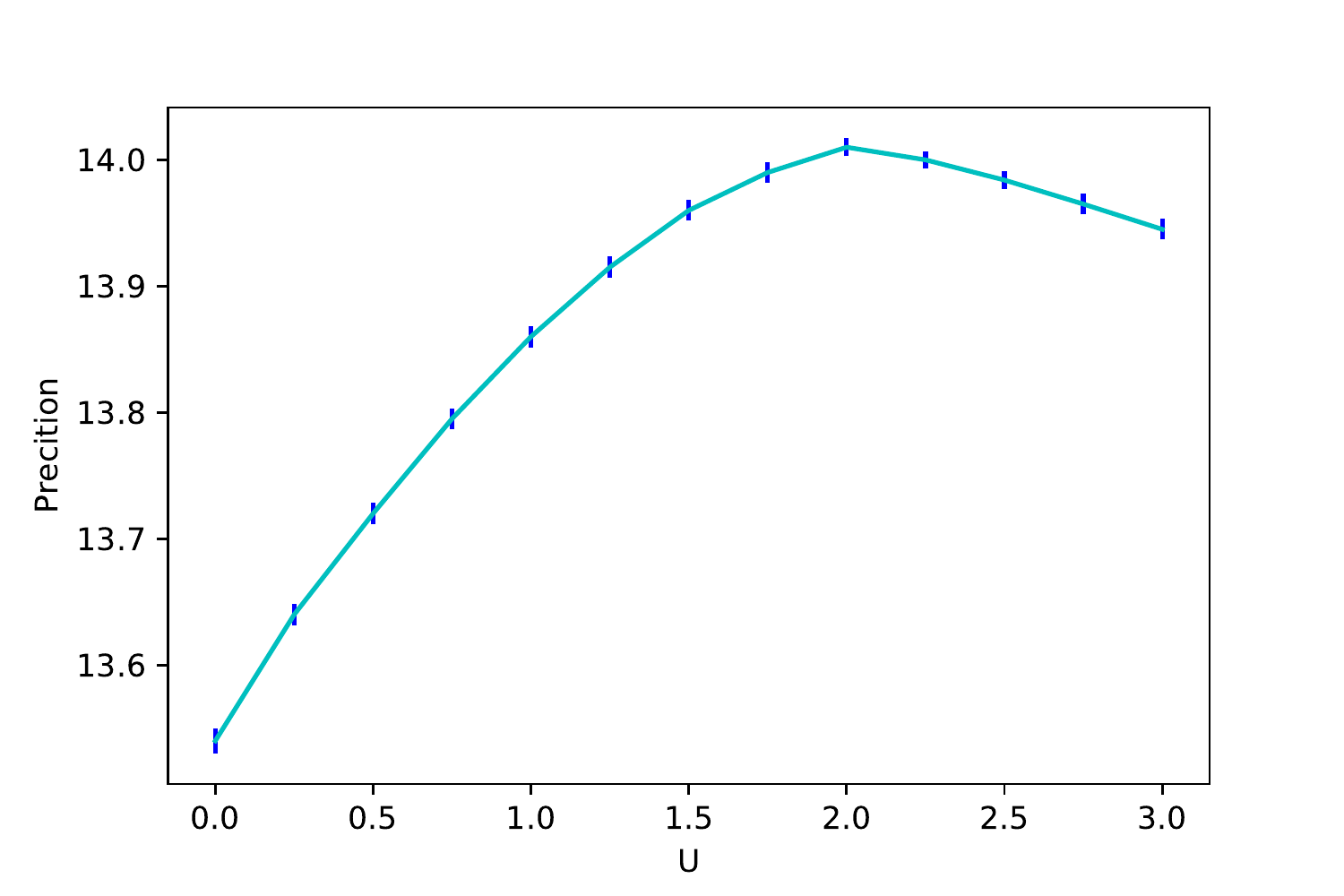}
\end{minipage}%
}%
\subfigure[impact of $U$ on diversity.]{
\begin{minipage}[t]{0.5\linewidth}
\centering
\includegraphics[width=4cm]{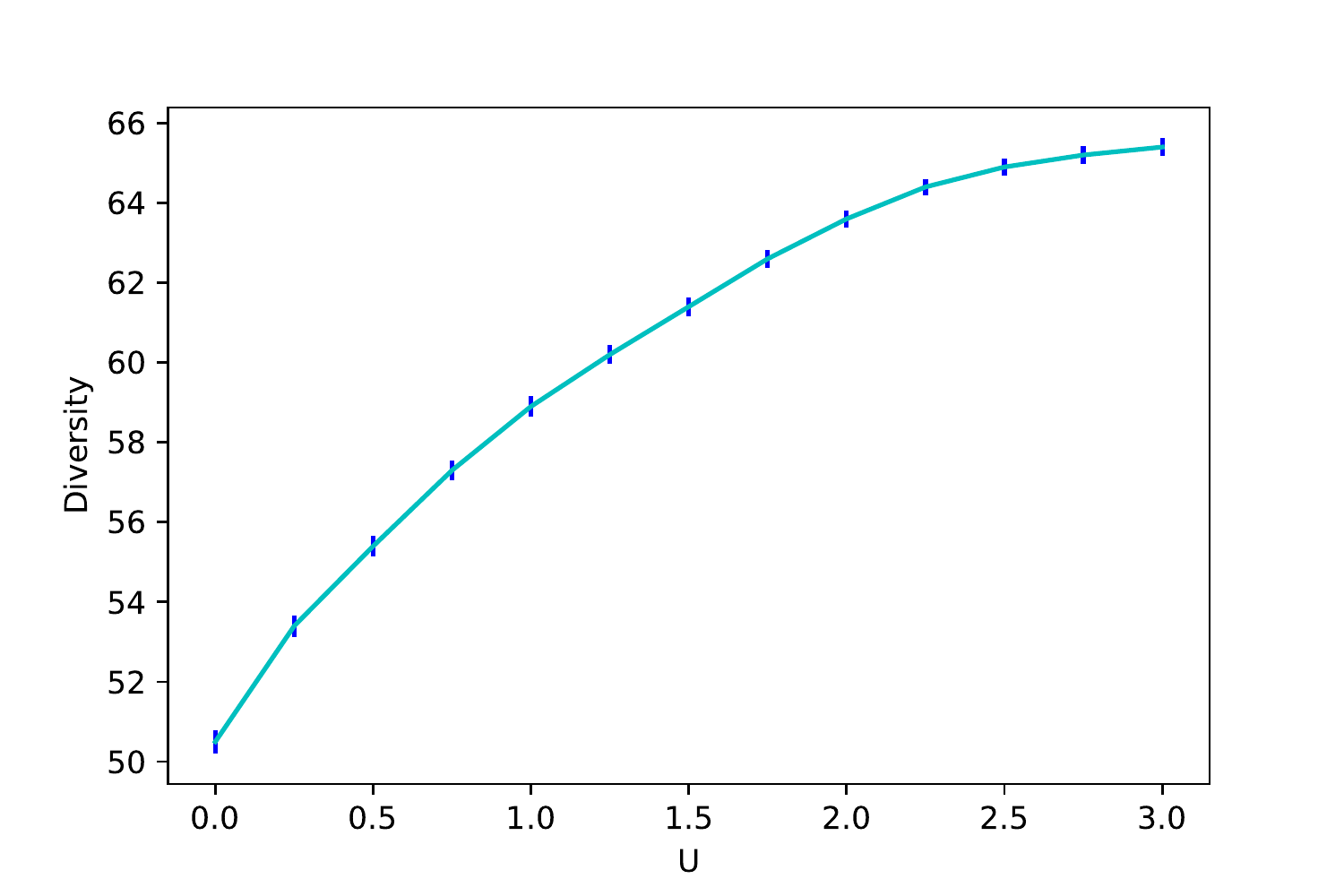}
\end{minipage}%
}%
\centering
\caption{Impact of trade-off parameter $U$.}
\label{fig:tradeoff}
\end{figure}

\begin{table*}[t]
\caption{Offline ${Precision@1}_{channel}$ \& ${Precision@2}_{channel}$ of Methods and Hyper-Parameters Investigation in DHANR model during a week.}
  \label{tab:freq2}
\centering
\begin{tabular}{|c|c|c|c|c|c|c|}
\hline

\multirow{2}{*}{method}& \multicolumn{3}{c|}{${Precision@1}_{channel}(\%)$} &\multicolumn{3}{c|}{${Precision@2}_{channel}2(\%)$}\\
\cline{2-7}
{}&channel 1&channel 2&channel 3&channel 1&channel 2&channel 3\\
\hline
UCI-AA&16.82&19.45&15.43&15.13&17.32&13.71\\
\cline{1-7}
UCI-AA-DLCM&17.34&20.27&16.32&15.31&17.52&13.88\\
\cline{1-7}
UCI-AA-PRM&17.96&20.54&16.56&15.33&17.51&13.90\\
\cline{1-7}
UCI-AA-DHANR($N_b=4$,$h=1$)&18.82&21.32&17.18&16.12&18.14&14.68\\
\cline{1-7}
Block($N_b=3$)&18.79&21.37&17.15&16.14&18.12&14.66\\
\cline{1-7}
Block($N_b=5$)&18.84&21.29&17.16&16.10&18.12&14.69\\
\cline{1-7}
Multiheads($h=2$)&18.85&21.28&17.21&16.13&18.13&14.67\\
\cline{1-7}
Multiheads($h=3$)&18.83&21.34&17.15&16.12&18.14&14.65\\
\hline
\end{tabular}
\end{table*}

\begin{table*}[t]
\caption{Performance improvements in online A/B test compared
to a DNN without re-ranking method during a week.}
  \label{tab:freq5}
\centering
\begin{tabular}{|c|c|c|c|c|c|c|c|c|}
\hline

\multirow{2}{*}{method}& \multicolumn{4}{c|}{CTR(\%)} &\multicolumn{4}{c|}{Diversity(\%)}\\
\cline{2-9}
{}&total&channel 1&channel 2&channel 3&total&channel 1&channel 2&channel 3\\
\hline
MMR&3.53&3.50&3.58&3.54&16.28&16.19&16.26&16.23\\
\cline{1-9}
MSD&3.58&3.59&3.55&3.61&15.31&15.27&15.36&15.32\\
\cline{1-9}
UCI-AA&5.62&5.60&5.66&5.64&26.23&26.21&26.24&26.25\\
\cline{1-9}
UCI-AA-DLCM&5.93&5.95&5.92&5.91&26.25&26.24&26.23&26.24\\
\cline{1-9}
UCI-AA-PRM&5.98&5.92&5.97&5.95&26.26&26.23&26.25&26.26\\
\cline{1-9}
UCI-AA-DHANR&6.26&6.29&6.24&6.23&26.26&26.25&26.23&26.25\\
\hline
\end{tabular}
\end{table*}

\subsection{Online Performance}
We conduct online experiments (A/B testing) on the E-commerce Recommendation online System.
The online performance of methods, measured in terms of CTR and diversity, is summarized in table 4.

Table 4 shows the performance improvements in online A/B test compared to a DNN model without re-ranking method. From table 4 we can conclude that by taking moderate amount of diversity into consideration, better performance can be achieved. All three methods, MMR, MSD and our UCI-AA model, increase the CTR and diversity. However, MMR and MSD do not improve the online metrics as much as our method. On average, our model outperforms DNN-based LTR without re-ranking models by 5.62\% and 26.23\% in terms of CTR and Diversity.
We can also conclude that re-ranking helps to increase the online metrics, as improvement can be observed regardless of the re-ranking methods. However, DLCM and PRM do not improve the online metrics as much as our method. Our model outperforms other algorithms in each channel.

\section{CONCLUSION}
In this paper, we propose a method to improve the precision and diversity for the homepage recommendation problem. Instead of optimizing evaluation metrics in a single channel, we take the homepage structure into consideration and design a two-stage architecture algorithm. During the first stage, we developed a general framework that employs linear programming to recommend items to proper channels under a whole class of diversity constraints. Then in the second stage, a DHANR model is proposed to refine the order of items in each channel. To verify the effectiveness of our method, we conduct offline and online experiments. Both the online and offline experiments demonstrate that our method could greatly improve the performance on real-world datasets.

Though we achieved our initial goal of increasing diversity and relevance of recommended items to users, there still remain many unexplored frontiers. One idea is to use Bandit algorithms to estimate the optimal number of items for each category. 

\clearpage
\bibliography{aaai22}

\end{document}